\documentclass[pra,aps,twocolumn,amsmath,amssymb,superscriptaddress,floatfix]{revtex4-2}
\usepackage{amsthm}
\usepackage{graphicx}
\usepackage{color}
\usepackage{comment}
\usepackage{amssymb}
\usepackage{graphicx}
\usepackage{eurosym}
\usepackage[utf8]{inputenc}
\usepackage[T1]{fontenc}
\usepackage{multirow}
\usepackage{xcolor}
\usepackage{hyperref}
\usepackage{algorithm}
\usepackage{algorithmic} 
\hypersetup{colorlinks, linkcolor=blue, citecolor=blue, urlcolor=blue}

\begin{document}
\title{Nonlocal effects via Local Quantum Fisher Information: Characterizations and Interpretations}



\author{R. Muthuganesan}
\email{Corresponding author: rajendramuthu@gmail.com}
\affiliation{Department of Physics and Nanotechnology, SRM Institute of Science and Technology, Kattankulathur--603203, Tamil Nadu, India}

%

\begin{abstract}
We introduce a quantum Fisher information based measurement-induced nonlocality (QFI-MIN), which quantifies the maximal statistical distinguishability induced by locally invariant unitary dynamics. The proposed measure inherits desirable properties including positivity, local unitary invariance, monotonicity under local operations, and immunity to the local ancilla problem. Analytical expressions are obtained for pure states, arbitrary two-qubit states, and two-qubit X states, revealing a direct connection with entanglement for pure systems. We further establish clear operational interpretations of QFI-MIN in quantum parameter estimation, local channel discrimination, and correlation-assisted communication. Its behavior under amplitude damping, depolarizing, and generalized amplitude damping channels demonstrates robustness against environmental noise. The proposed framework provides a physically consistent and operationally meaningful quantifier of quantum correlations, linking nonlocality, quantum metrology, and quantum information processing. 
\end{abstract}

\maketitle

\section{Introduction}
\label{sec:1}
Composite quantum systems exhibit many intrinsic features that have no classical counterpart, among which nonclassical correlations constitute one of the most fundamental aspects \cite{Nielsen2000,Horodecki2009}. These correlations serve as essential resources for various quantum information processing tasks, including quantum teleportation, dense coding, and quantum cryptography. In this context, quantifying nonclassical correlations in composite quantum systems, particularly in bipartite systems, has become a central problem in quantum information theory. Beyond entanglement, several measures of quantum correlations have been developed \cite{Ollivier2001,Henderson2001,Dakic2010,Luo2008}. In particular, measurement-based approaches play an important role in quantifying correlations in composite systems. Some notable measures include quantum discord \cite{Ollivier2001,Henderson2001}, measurement-induced disturbance (MID) \cite{Luo2008}, uncertainty-induced nonlocality (UIN) \cite{Wu2014}, and measurement-induced nonlocality (MIN) \cite{LUOMIN2011,Muthuganesan2017,Muthuganesan2020}. Among these, MIN captures the maximal global effect induced by locally invariant measurements and provides a distinct perspective on nonlocal features that are not necessarily revealed by entanglement alone.

On the other hand, quantum Fisher information (QFI) plays a fundamental role in quantum information theory, particularly in quantum estimation, statistical distinguishability, and information geometry \cite{Helstrom1976,Braunstein1994,Paris2009,Giovannetti2011}. As a central quantity governing the ultimate precision limits of parameter estimation, QFI provides a natural metric on the space of quantum states and has been extensively used to characterize quantum resources in quantum metrology and related areas. Recent studies have also connected local QFI with measures of quantum correlations, particularly discord-type correlations. Moreover, QFI has proven to be a powerful tool for quantifying various quantum resources, including nonclassical correlations and quantum coherence, and has been widely employed for entanglement detection and characterization \cite{Modi2011,Toth2014,Girolami2014,Li2013}.

Recent studies have established close connections between QFI and quantum correlations, demonstrating that Fisher-information-based quantities can capture nonclassical features relevant for quantum metrology and information processing \cite{Modi2011,Girolami2014}. In this work, we introduce a new measure of measurement-induced nonlocality based on quantum Fisher information, termed QFI-based MIN. Unlike geometric formulations that quantify finite state disturbance under local measurements, the proposed measure captures the maximal infinitesimal statistical distinguishability induced by locally invariant unitary dynamics. In this sense, it unifies the disturbance-based perspective of MIN with the information-geometric framework of quantum Fisher information.

The proposed measure admits a direct operational interpretation. Specifically, it quantifies the maximum sensitivity of a bipartite quantum state to locally invariant unitary encodings on one subsystem, thereby identifying the optimal distinguishability rate for such encodings. This connects measurement-induced nonlocality with fundamental tasks in quantum information theory, including channel discrimination, parameter estimation, and classical communication under constrained encoding. In particular, we show that QFI-based MIN determines the leading-order behavior of the Holevo information in locally invariant communication schemes, establishing it as a measure of correlation-assisted communication power in the small-signal regime.

From a structural perspective, QFI-based MIN naturally extends and strengthens existing MIN-type measures. It can be viewed as a Fisher-information generalization of skew-information-based nonlocality measures, while simultaneously providing a physically consistent alternative to geometric MIN formulations. Furthermore, the measure inherits key desirable properties, including invariance under local unitary transformations and robustness under the addition of uncorrelated ancillas, thereby resolving limitations of earlier formulations.

We develop a comprehensive theoretical framework for QFI-based MIN. We establish its fundamental properties, including positivity, invariance, and behavior under local operations. For pure states, we derive closed-form expressions and show a direct connection with entanglement measures. For mixed states, we obtain explicit formulas for important classes such as two-qubit X states. In addition, we present a general formulation for arbitrary two-qubit states in Bloch representation, revealing a natural branch structure depending on the degeneracy of the reduced subsystem. Overall, the present work provides a unified and operationally meaningful extension of measurement-induced nonlocality, bridging geometric, information-theoretic, and metrological perspectives. These results not only deepen the understanding of quantum correlations but also highlight their role as resources in fundamental information-processing tasks.

The remainder of this paper is organized as follows. In Sec.~II, we briefly review the concept of quantum Fisher information and its fundamental properties. Section~\ref{QFI_MIN} introduces the proposed quantum Fisher information based measurement-induced nonlocality (QFI-MIN) and establishes its basic properties. In Sec.~\ref{Pure}, explicit analytical expressions are derived for bipartite pure states, highlighting the connection between QFI-MIN and entanglement. Section~\ref{Ancilla} discusses the local ancilla problem and demonstrates the robustness of QFI-MIN against trivial ancilla extensions. The operational interpretations of QFI-MIN  are presented in Sec.~\ref{inter}. A general formulation for arbitrary two-qubit states and analytical results for two-qubit X states are obtained in Sec.~\ref{Two-Qubit}. In Sec.~\ref{Dynamics}, we investigate the behavior of QFI-MIN under various noisy quantum channels and compare its robustness with other MIN measures. Finally, the main conclusions are summarized.

\section{Quantum Fisher Information}
\label{QFI}
Quantum Fisher Information (QFI) is one of the fundamental quantities in quantum estimation theory and quantum metrology, as it characterizes the ultimate precision attainable in parameter-estimation protocols \cite{Helstrom1976,Braunstein1994,Paris2009}. Beyond its metrological significance, QFI has emerged as an important tool for quantifying various quantum resources, including coherence, asymmetry, entanglement, and nonclassical correlations \cite{Girolami2014,Toth2014,Modi2011,Li2013}.

Consider a quantum state $\rho_{\theta}$ that depends on an unknown parameter $\theta$. The parameter encoding is achieved through a unitary transformation generated by an observable $K$,
\begin{equation}
\rho_{\theta}=U_{\theta}\rho U_{\theta}^{\dagger},
\qquad
U_{\theta}=e^{-iK\theta},
\end{equation}
where $\rho$ denotes the initial probe state. The quantum Fisher information associated with the parameter $\theta$ is defined as
\begin{equation}
F(\rho_{\theta})
=
\frac{1}{4}
\mathrm{Tr}
\left(
\rho_{\theta}L_{\theta}^{2}
\right),
\end{equation}
where $L_{\theta}$ is the symmetric logarithmic derivative (SLD), implicitly determined by
\begin{equation}
\frac{d\rho_{\theta}}{d\theta}
=
\frac{1}{2}
\left(
\rho_{\theta}L_{\theta}
+
L_{\theta}\rho_{\theta}
\right).
\end{equation}

Let the spectral decomposition of the density operator be
\begin{equation}
\rho
=
\sum_i p_i |\psi_i\rangle\langle\psi_i|,
\end{equation}
with $p_i\ge0$ and $\sum_i p_i=1$. The QFI corresponding to the generator $K$ can then be expressed as \cite{Braunstein1994,Toth2014}
\begin{equation}
F(\rho,K)
=
\frac{1}{2}
\sum_{i\neq j}
\frac{(p_i-p_j)^2}{p_i+p_j}
\left|
\langle\psi_i|K|\psi_j\rangle
\right|^2.
\label{QFI}
\end{equation}

The quantum Fisher information satisfies several well-established information-theoretic properties \cite{Braunstein1994,Paris2009,Toth2014}. The quantum Fisher information possesses several important properties:

\begin{enumerate}
\item \textbf{Positivity:}
\begin{equation}
F(\rho,K)\ge0.
\end{equation}

\item \textbf{Pure-state reduction:}
For a pure state $\rho=|\psi\rangle\langle\psi|$, the QFI reduces to the variance $\Delta K$ of the generator,
\begin{equation}
F(\rho,K)
=
(\Delta K)^2.
\end{equation}
For mixed states,
\begin{equation}
0\le F(\rho,K)\le (\Delta K)^2.
\end{equation}

\item \textbf{Unitary invariance:}
If a unitary operator $U$ commutes with $K$, then
\begin{equation}
F(U\rho U^{\dagger},K)
=
F(\rho,K).
\end{equation}

\item \textbf{Convexity:}
For an ensemble of quantum states ${\rho_i}$ with probabilities ${\lambda_i}$,
\begin{equation}
F\left(
\sum_i \lambda_i\rho_i,K
\right)
\le
\sum_i \lambda_i F(\rho_i,K),
\end{equation}
where $\lambda_i\ge0$ and $\sum_i\lambda_i=1$.

\item \textbf{Basis independence:}
The value of QFI is independent of the choice of orthonormal basis used to represent the density matrix.

\item \textbf{Superadditivity:}
For a bipartite state $\rho_{AB}$ and local observables $K_A$ and $K_B$,
\begin{equation}
F\left(
\rho_{AB},
K_A\otimes I_B + I_A\otimes K_B
\right)
\ge
F(\rho_A,K_A)
+
F(\rho_B,K_B),
\end{equation}
with equality holding for product states.
\end{enumerate}

Owing to these properties, QFI serves not only as a central quantity in quantum metrology but also as a powerful indicator of quantum resources and nonclassical features in composite quantum systems.

\section{Quantum Fisher Information Based MIN}
\label{QFI_MIN}
Measurement-induced nonlocality (MIN) was originally introduced as the maximal global disturbance caused by locally invariant measurements on one subsystem~\cite{LUOMIN2011}. Later refinements replaced the Hilbert--Schmidt norm with information-theoretic quantities such as skew information, leading to uncertainty-induced nonlocality (UIN)~\cite{Wu2014}. Motivated by the metrological significance of quantum Fisher information (QFI), we define here a QFI-based variant of MIN by preserving the local invariance constraint while replacing the disturbance functional with QFI.

Let $\rho_{AB}$ be a bipartite quantum state acting on $\mathcal{H}_A \otimes \mathcal{H}_B$, and  $\rho_{A (B)}= \mathrm{Tr}_{B(A)}(\rho_{AB})$ be the marginal state in the Hilbert space $\mathcal{H}_{A(B)}$ . The local quantum Fisher information based measurement-induced nonlocality (QFI-MIN) with respect to subsystem $A$ is defined as
\begin{equation}
\mathcal{N}_F(\rho_{AB})
:=
\sup_{K_A \in \mathcal{L}_A^{\mathrm{inv}}}
F_Q\!\left(\rho_{AB},\, K_A \otimes \mathbb{I}_B \right),
\end{equation}
where $F_Q(\rho,H)$ denotes the quantum Fisher information of $\rho$ with respect to the Hermitian operator $H$, and
\begin{equation}
\mathcal{L}_A^{\mathrm{inv}}
=
\left\{
K_A = K_A^\dagger :
[K_A,\rho_A] = 0,\;
\mathrm{Tr}(K_A^2)=1
\right\}.
\end{equation}

The constraint $[K_A,\rho_A]=0$ ensures that the generator does not disturb the local marginal, in analogy with the locally invariant measurement condition in the original MIN framework~\cite{LUOMIN2011}.
The quantity $\mathcal{N}_F(\rho_{AB})$ satisfies the following properties:

\begin{enumerate}
    \item \textbf{Positivity} $\mathcal{N}_F^A(\rho_{AB}) \ge 0$;
    \item \textbf{Local unitary invariance}
    \begin{equation}
    \mathcal{N}_F\!\left((U_A \otimes U_B)\rho_{AB}(U_A^\dagger \otimes U_B^\dagger)\right)
    =
    \mathcal{N}_F(\rho_{AB});
    \end{equation}
    \item \textbf{Monotonicity under local CPTP maps on $B$}
    \begin{equation}
    \mathcal{N}_F\!\left((\mathbb{I}_A \otimes \Lambda_B)(\rho_{AB})\right)
    \le
    \mathcal{N}_F(\rho_{AB});
    \end{equation}
    \item \textbf{Vanishing condition} If $\rho_{AB} = \rho_A \otimes \rho_B$ or if $\rho_{AB}$ is a classical--quantum state with nondegenerate $\rho_A$, then
    \begin{equation}
    \mathcal{N}_F(\rho_{AB}) = 0;
    \end{equation}
    \item \textbf{Pure-state reduction} For a pure state $|\psi\rangle_{AB}$,
    \begin{equation}
    \mathcal{N}_F(|\psi\rangle)
    =
    4 \sup_{K_A \in \mathcal{L}_A^{\mathrm{inv}}}
    \mathrm{Var}_{|\psi\rangle}(K_A \otimes \mathbb{I}_B).
    \end{equation}
\end{enumerate}

Here, we sketch the proof of  the above-mentioned properties. 
\begin{proof}
\textbf{(1) Positivity.}
Quantum Fisher information satisfies $F_Q(\rho,H) \ge 0$ for all $\rho$ and $H$. Hence the supremum is nonnegative.

\textbf{(2) Local unitary invariance.}
Let $\rho'_{AB} = (U_A \otimes U_B)\rho_{AB}(U_A^\dagger \otimes U_B^\dagger)$ and $\rho_A' = U_A \rho_A U_A^\dagger$. If $K_A \in \mathcal{L}_A^{\mathrm{inv}}$, then $K_A' = U_A K_A U_A^\dagger$ satisfies $[K_A',\rho_A'] = 0$ and $\mathrm{Tr}[(K_A')^2] = 1$. By unitary covariance of QFI,
\begin{equation}
F_Q(\rho'_{AB}, K_A' \otimes \mathbb{I}_B)
=
F_Q(\rho_{AB}, K_A \otimes \mathbb{I}_B).
\end{equation}
Taking the supremum over admissible generators yields the result.

\textbf{(3) Monotonicity.}
Let $\Lambda_B$ be a CPTP map acting on subsystem $B$. Since $\rho_A$ remains unchanged, the admissible set $\mathcal{L}_A^{\mathrm{inv}}$ is unaffected. By monotonicity of QFI under CPTP maps,
\begin{equation}
F_Q\!\left((\mathbb{I}_A \otimes \Lambda_B)(\rho_{AB}), K_A \otimes \mathbb{I}_B\right)
\le
F_Q(\rho_{AB}, K_A \otimes \mathbb{I}_B).
\end{equation}
Taking the supremum proves the claim.

\textbf{(4) Vanishing condition.}
If $\rho_{AB} = \rho_A \otimes \rho_B$ and $\rho_A$ is nondegenerate, then any $K_A$ satisfying $[K_A,\rho_A]=0$ is diagonal in the eigenbasis of $\rho_A$, implying
\begin{equation}
[K_A \otimes \mathbb{I}_B, \rho_{AB}] = 0.
\end{equation}
Thus $F_Q(\rho_{AB}, K_A \otimes \mathbb{I}_B) = 0$, and the supremum vanishes. The same argument applies to classical--quantum states.

\textbf{(5) Pure-state reduction.}
For pure states, the quantum Fisher information reduces to
\begin{equation}
F_Q(|\psi\rangle, H) = 4\,\mathrm{Var}_{|\psi\rangle}(H).
\end{equation}
Substituting $H = K_A \otimes \mathbb{I}_B$ yields the result.
\end{proof}

The quantity $\mathcal{N}_F(|\psi\rangle)$ should be interpreted as a QFI-based generalization of measurement-induced nonlocality rather than the original MIN. It inherits the local invariance principle while incorporating the contractive and operationally meaningful properties of quantum Fisher information. Care must be taken when $\rho_A$ is degenerate, as the admissible generator set becomes larger and may lead to nonzero values even for classical--quantum states.

\section{Explicit expression for pure states}\label{Pure}
Let
\begin{equation}
|\psi\rangle_{AB}=\sum_{i=1}^r \sqrt{\lambda_i}\,|i\rangle_A|i\rangle_B
\end{equation}
be the Schmidt decomposition of a bipartite pure state, where $\lambda_i\ge 0$ and $\sum_i\lambda_i=1$. Then the reduced state is $\rho_A=\sum_i \lambda_i |i\rangle\langle i|.$ The Schmidt decomposition provides a complete characterization of bipartite pure-state entanglement \cite{Nielsen2000,Horodecki2009}. Since for pure states the quantum Fisher information reduces to four times the variance i.e.,  \cite{Braunstein1994,Toth2014}
\begin{equation}
\mathcal N_F(|\psi\rangle)
= 4\sup_{K_A}
\mathrm{Var}_{|\psi\rangle}(K_A\otimes\mathbb I_B). \nonumber
\end{equation}
Because $[K_A,\rho_A]=0$, the Hermitian operator $K_A$ is diagonal in the Schmidt basis,
\begin{equation}
K_A=\sum_i k_i |i\rangle\langle i|,
\qquad
\sum_i k_i^2=1. \nonumber
\end{equation}
Therefore,
\begin{equation}
\langle\psi|K_A\otimes\mathbb I_B|\psi\rangle
=
\sum_i \lambda_i k_i, \nonumber
\end{equation}
and
\begin{equation}
\langle\psi|K_A^2\otimes\mathbb I_B|\psi\rangle
=
\sum_i \lambda_i k_i^2. \nonumber
\end{equation}
Hence,
\begin{equation}
\mathcal N_F^A(|\psi\rangle)
=
4\sup_{\sum_i k_i^2=1}
\left[
\sum_i \lambda_i k_i^2-\left(\sum_i \lambda_i k_i\right)^2
\right]. \nonumber
\end{equation}
Writing $\mathbf{k}=(k_1,\dots,k_r)^T$, $\lambda=(\lambda_1,\dots,\lambda_r)^T$, and $D=\mathrm{diag}(\lambda_1,\dots,\lambda_r)$, the above expression becomes
\begin{equation}
\mathcal N_F(|\psi\rangle)
=
4\sup_{\mathbf{k}^T\mathbf{k}=1}
\mathbf{k}^T(D-\lambda\lambda^T)\mathbf{k}. \nonumber
\end{equation}
The optimization of quadratic forms over normalized vectors reduces to the largest eigenvalue problem of the associated symmetric matrix \cite{HornJohnson}.
\begin{equation}
\mathcal N_F(|\psi\rangle)
=
4\,\lambda_{\max}(D-\lambda\lambda^T)
\end{equation}
where $\lambda_{\max}$ denotes the largest eigenvalue.

For a two-qubit pure state
\begin{equation}
|\psi\rangle=\sqrt{\lambda_1}|00\rangle+\sqrt{\lambda_2}|11\rangle,
\qquad
\lambda_1+\lambda_2=1, \nonumber
\end{equation}
one obtains
\begin{equation}
\mathcal N_F(|\psi\rangle)=8\lambda_1\lambda_2.
\end{equation}
Equivalently, in terms of concurrence $C=2\sqrt{\lambda_1\lambda_2}$,
\begin{equation}
\mathcal N_F^A(|\psi\rangle)=2C^2.
\end{equation}
For pure states, the QFI-MIN is exactly the maximum locally invariant variance, and for two-qubit system, it is directly proportional to the square of concurrence \cite{Wootters1998}. This relation establishes a direct connection between QFI-MIN and bipartite entanglement for pure states, similar to several Fisher-information-based quantum correlation measures reported in the literature \cite{Modi2011,Girolami2013}.

\section{Local Ancilla Problem}
\label{Ancilla}
For a given bipartite state $\rho_{AB}$, let $\sigma_C$ be an arbitrary state on an auxiliary system $C$. The addition of a local ancilla demonstrates that Hilbert–Schmidt norm-based measurement-induced nonlocality (HS-MIN) is not a bona fide measure of quantum correlations, owing to the non-contractive nature of the Hilbert–Schmidt norm \cite{Piani2012}. To address this issue, several alternative versions of MIN have been proposed and investigated, each designed to satisfy the fundamental requirements of a valid quantum correlation measure \cite{Muthuganesan2017,Muthuganesan2020,MUTHUGANESAN2023129250,BHUVANESWARI2025130939,Hu2012,Ciccarello2014,Girolami2013}.

After the addition of a local ancilla state $\sigma_C$, for any admissible generator $K_A$ satisfying $[K_A,\rho_A]=0$, the local quantum Fisher information of the extended state satisfies
\begin{equation}
F_Q(\rho_{AB}\otimes \sigma_C, K_A\otimes I_{BC})
= F_Q(\rho_{AB}, K_A\otimes I_B),
\end{equation}
which implies that the quantum Fisher information remains invariant under tensoring an independent ancilla when the generator acts solely on the original subsystem. Consequently, by taking the supremum over all admissible generators $K_A$, the QFI-based measurement-induced nonlocality obeys
\begin{equation}
\mathcal N_F(\rho_{AB}\otimes \sigma_C)
=
\mathcal N_F(\rho_{AB}). \nonumber
\end{equation}

 Therefore, unlike Hilbert–Schmidt MIN, the proposed QFI-based MIN is immune to the local ancilla problem. Owing to the invariance of quantum Fisher information under the addition of an independent ancilla, QFI-based MIN remains unchanged and thus constitutes a physically consistent measure of quantum correlations.

\section{Operational Interpretations of QFI-Based MIN}
\label{inter}
One of the main advantages of the proposed QFI-based measurement-induced nonlocality (QFI-MIN) is that it admits direct operational interpretations in several fundamental quantum-information tasks. Since quantum Fisher information governs the ultimate distinguishability of nearby quantum states, the quantity $\mathcal{N}_F(\rho_{AB})$ naturally characterizes the maximum information-processing capability that can be extracted from locally invariant encodings. In this section, we establish its connections with local channel discrimination, quantum metrology, and correlation-assisted communication.

\subsection{Local Channel Discrimination}

Consider the family of locally invariant unitary channels

\begin{equation}
\mathcal{E}^{(K)}_{\theta}(\rho)
=
\left(e^{-i\theta K_A}\otimes I_B\right)
\rho
\left(e^{i\theta K_A}\otimes I_B\right),
\end{equation}
where the generator $K_A$ satisfies
\begin{equation}
[K_A,\rho_A]=0,
\qquad
\mathrm{Tr}(K_A^2)=1. \nonumber
\end{equation}
For sufficiently small $\theta$, the distinguishability between
$\rho_{AB}$ and
$\mathcal{E}^{(K)}_{\theta}(\rho_{AB})$
is quantified by the Bures distance,
\begin{equation}
F_Q(\rho_{AB},K_A\otimes I_B)
=
4
\lim_{\theta\rightarrow0}
\frac{
d_B^2
\!\left(
\rho_{AB},
\mathcal{E}^{(K)}_{\theta}(\rho_{AB})
\right)
}
{\theta^2}.
\end{equation}

Consequently,

\begin{equation}
\mathcal{N}_F(\rho_{AB})
=
4
\sup_{K_A\in\mathcal{L}^{\rm inv}_A}
\lim_{\theta\rightarrow0}
\frac{
d_B^2
\!\left(
\rho_{AB},
\mathcal{E}^{(K)}_{\theta}(\rho_{AB})
\right)
}
{\theta^2}.
\end{equation}
The relation between quantum Fisher information, Bures distance, and infinitesimal channel distinguishability has been extensively studied in quantum estimation theory and quantum channel discrimination \cite{Braunstein1994,Paris2009,Helstrom1976}. Therefore, QFI-MIN quantifies the maximum infinitesimal distinguishability achievable through locally invariant unitary perturbations. In this sense, it represents the optimal sensitivity of the state $\rho_{AB}$ for discriminating nearby local quantum channels that leave the reduced state $\rho_A$ unchanged.

\subsection{Quantum Metrology and Parameter Estimation}

Let a parameter $\theta$ be encoded through the locally invariant unitary transformation

\begin{equation}
\rho_{AB}(\theta)
=
\left(e^{-i\theta K_A}\otimes I_B\right)
\rho_{AB}
\left(e^{i\theta K_A}\otimes I_B\right), \nonumber
\end{equation}

with $[K_A,\rho_A]=0$.

According to the quantum Cram\'er--Rao theorem, any unbiased estimator $\hat{\theta}$ satisfies

\begin{equation}
\mathrm{Var}(\hat{\theta})
\ge
\frac{1}
     {F_Q(\rho_{AB},K_A\otimes I_B)}.
\end{equation}

Optimizing over all admissible generators yields

\begin{equation}
\mathrm{Var}(\hat{\theta})
\ge
\frac{1}
     {\mathcal{N}_F(\rho_{AB})}.
\end{equation}
The quantum Cram\'er-Rao bound establishes the fundamental precision limit in parameter estimation and identifies quantum Fisher information as the relevant metrological resource \cite{Braunstein1994,Paris2009,Giovannetti2011}.
Hence, QFI-MIN determines the ultimate precision limit for estimating parameters encoded through locally invariant dynamics. Larger values of QFI-MIN correspond to greater metrological usefulness and enhanced estimation sensitivity.

\subsection{Standard Quantum Limit and Heisenberg Scaling}

Consider $n$ independent copies of the probe state,
\begin{equation}
\rho_{AB}^{\otimes n},
\end{equation}
with collective generator
\begin{equation}
H_n
=
\sum_{m=1}^{n}
K_A^{(m)}
\otimes
I_B^{(m)}.
\end{equation}
Since quantum Fisher information is additive for independent probes,
\begin{equation}
F_Q
\!\left(
\rho_{AB}^{\otimes n},
H_n
\right)
=
n\,
F_Q(\rho_{AB},K_A\otimes I_B).
\end{equation}
Therefore,
\begin{equation}
\mathcal{N}_{F}^{n}
=
n\,\mathcal{N}_F(\rho_{AB}),
\end{equation}
and the corresponding estimation precision scales as
\begin{equation}
\mathrm{Var}(\hat{\theta})
\ge
\frac{1} {n\,\mathcal{N}_F(\rho_{AB})},
\end{equation}
which corresponds to the standard quantum limit (SQL).

A stronger Heisenberg scaling,
\begin{equation}
\mathrm{Var}(\hat{\theta})
\sim
\frac{1}{n^2},
\end{equation}
can arise only when collective entangled probes and collective locally invariant generators are employed. The distinction between the standard quantum limit and Heisenberg scaling is one of the central themes of quantum metrology and arises from the use of independent and entangled probe states, respectively \cite{Giovannetti2004,Giovannetti2011,Toth2014}. Thus, QFI-MIN identifies the locally invariant metrological resource responsible for determining the attainable scaling law.

\subsection{Correlation-Assisted Classical Communication}

Suppose classical information is encoded through locally invariant unitary modulations

\begin{equation}
\rho_{AB}^{(x)}
=
\left(
e^{-i\theta s_x K_A}
\otimes I_B
\right)
\rho_{AB}
\left(
e^{i\theta s_x K_A}
\otimes I_B
\right), \nonumber
\end{equation}
where $\{s_x\}$ denotes the signal alphabet with $\sum_x p_x s_x =0$. For weak modulation amplitudes ($\theta\ll1$), the Holevo information admits the expansion
\begin{equation}
\chi
=
\frac{\theta^2}
     {8\ln 2}
\,p(s)\,
F_Q(\rho_{AB},K_A\otimes I_B)
+
O(\theta^2),
\end{equation}
where
\begin{equation}
p(s)
=
\sum_x p_x s_x^2
-
\left(
\sum_x p_x s_x
\right)^2.
\end{equation}
Optimizing over all admissible generators gives
\begin{equation}
\chi_{\max}
=
\frac{\theta^2}
     {8\ln 2}
\,p(s)\,
\mathcal{N}_F(\rho_{AB})
+
O(\theta^2).
\end{equation}
The Holevo quantity provides the ultimate upper bound on the accessible classical information transmitted through quantum states and plays a central role in quantum communication theory \cite{Holevo1973,Holevo1998,Schumacher1997}.
Therefore, QFI-MIN determines the leading-order communication rate achievable under locally invariant encodings. Since the reduced state $\rho_A$ remains unchanged for all codewords, the transmitted information originates entirely from bipartite quantum correlations. In this sense, QFI-MIN quantifies the correlation-assisted communication capability of the state.

The above results reveal that QFI-MIN possesses a unified operational significance. Therefore, unlike geometric formulations of measurement-induced nonlocality, the proposed QFI-MIN is directly linked to experimentally meaningful information-processing tasks. It provides an operationally motivated measure of quantum correlations that unifies channel discrimination, quantum metrology, and communication theory within a single information-geometric framework.

\section{General formulation for arbitrary two-qubit states}
\label{Two-Qubit}
Any two-qubit state can be written in the Bloch form
\begin{equation}
\rho=\frac14\Bigl(
I\otimes I+\vec x \cdot \vec\sigma\otimes I+I\otimes \vec y\cdot \vec\sigma
+\sum_{i,j=1}^3 T_{ij}\,\sigma_i\otimes \sigma_j
\Bigr),
\label{eq:bloch2qubit}
\end{equation}
where $\vec x,\vec y\in\mathbb R^3$ are the local Bloch vectors and $T=(T_{ij})$ is the real correlation matrix. The reduced density matrix on subsystem $A$ is
\begin{equation}
\rho_A=\frac12\bigl(I+\vec x\cdot \vec\sigma\bigr).
\label{eq:rhoA_bloch}
\end{equation}

Let
\begin{equation}
K_A(\vec n)=\frac{1}{\sqrt2}\,\vec n\cdot \vec\sigma,
\qquad |\vec n|=1.
\label{eq:KA_n}
\end{equation}
Then $[K_A(\vec n)^2]=1$, and the local invariance condition
\begin{equation}
[K_A(\vec n),\rho_A]=0 \nonumber
\label{eq:comm_constraint}
\end{equation}
is equivalent to $\vec n\times \vec x=0$. Define the real symmetric matrix $M(\rho)$ with entries
\begin{equation}
M_{ij}(\rho)
:=
\sum_{m,n:\,\lambda_m+\lambda_n>0}
\frac{(\lambda_m-\lambda_n)^2}{\lambda_m+\lambda_n}
\langle m|\sigma_i\otimes I|n\rangle
\langle n|\sigma_j\otimes I|m\rangle, \nonumber
\label{eq:Mmatrix}
\end{equation}
with $\rho=\sum_m \lambda_m |m\rangle\langle m|$.

For any two-qubit state $\rho$, the QFI-based measurement-induced nonlocality with respect to subsystem $A$ is given by
\begin{equation}
\mathcal N_F(\rho)=
\begin{cases}
\hat x^{\,T} M(\rho)\,\hat x, & \vec x\neq 0,\\[6pt]
\lambda_{\max}(M(\rho)), & \vec x=0,
\end{cases}
\label{eq:general_2qubit_formula}
\end{equation}
where $\hat x=\vec x/|\vec x|$ and $\lambda_{\max}(M)$ denotes the largest eigenvalue of $M$.

\begin{proof}
Using the spectral formula for the quantum Fisher information,
\begin{equation}
F_Q(\rho,H)=
2\sum_{m,n:\,\lambda_m+\lambda_n>0}
\frac{(\lambda_m-\lambda_n)^2}{\lambda_m+\lambda_n}
|\langle m|H|n\rangle|^2, \nonumber
\label{eq:qfi_spectral}
\end{equation}
and substituting
\begin{equation}
H=K_A(\vec n)\otimes I
=\frac{1}{\sqrt2}\sum_{i=1}^3 n_i\,\sigma_i\otimes I, \nonumber
\end{equation}
one obtains
\begin{equation}
F_Q\!\left(\rho,\frac{\vec n\cdot \vec\sigma}{\sqrt2}\otimes I\right)
=
\vec n^{\,T}M(\rho)\,\vec n. \nonumber
\label{eq:qfi_quadratic_form}
\end{equation}
If $\vec x\neq 0$, the constraint $[K_A(\vec n),\rho_A]=0$ implies $\vec n=\pm \hat x$, hence
\begin{equation}
\mathcal N_F(\rho)=\hat x^{\,T}M(\rho)\,\hat x. \nonumber
\end{equation}
If $\vec x=0$, every direction $\vec n$ is admissible, and maximizing the quadratic form over all unit vectors yields
\begin{equation}
\mathcal N_F(\rho)=\lambda_{\max}(M(\rho)).
\end{equation}
This completes the proof.
\end{proof}

\subsection{Two-qubit X states}
\label{Two-QubitX}
Consider a two-qubit X state
\begin{equation}
\rho_X=
\begin{pmatrix}
a & 0 & 0 & u\\
0 & b & v & 0\\
0 & v^* & c & 0\\
u^* & 0 & 0 & d
\end{pmatrix},
\qquad
a,b,c,d\ge 0,\qquad
a+b+c+d=1,
\label{eq:Xstate} \nonumber
\end{equation}
with positivity conditions $|u|^2\le ad$ and $|v|^2\le bc$.
Its reduced density matrix on subsystem $A$ is
\begin{equation}
\rho_A=\text{Tr}_B(\rho_X)=
\begin{pmatrix}
a+b & 0\\
0 & c+d
\end{pmatrix}.
\label{eq:rhoA_X} 
\end{equation}

If $a+b\neq c+d$, i.e. $\rho_A$ is nondegenerate, then the QFI-based measurement-induced nonlocality with respect to subsystem $A$ is
\begin{equation}
\mathcal N_F(\rho_X)
=
8\left(
\frac{|u|^2}{a+d}
+
\frac{|v|^2}{b+c}
\right).
\label{eq:Xstate_main_result}
\end{equation}

\begin{proof}
Since $\rho_A$ is diagonal and nondegenerate, the condition $[K_A,\rho_A]=0$ implies that every admissible local generator is diagonal:
\begin{equation}
K_A=\begin{pmatrix}k_1&0\\0&k_2\end{pmatrix}, 
\qquad
k_1^2+k_2^2=1. \nonumber
\label{eq:KA_X}
\end{equation}
Hence,
\begin{equation}
K_A\otimes \mathbb I_B
=
\mathrm{diag}(k_1,k_1,k_2,k_2). \nonumber
\end{equation}
Write
\begin{equation}
K_A=\alpha \mathbb I+\beta \sigma_z,
\qquad
\alpha=\frac{k_1+k_2}{2},\qquad
\beta=\frac{k_1-k_2}{2}. \nonumber
\end{equation}
Because adding a multiple of the identity does not affect the QFI,
\begin{equation}
F_Q(\rho_X,K_A\otimes\mathbb I_B)
=
F_Q(\rho_X,\beta\,\sigma_z\otimes\mathbb I_B). \nonumber
\end{equation}
Under the constraint $k_1^2+k_2^2=1$, the quantity $(k_1-k_2)^2$ is maximized by $k_{1,2}=\pm\frac{1}{\sqrt2}$, so the optimal generator is
\begin{equation}
K_A^{\mathrm{opt}}=\frac{\sigma_z}{\sqrt2}. \nonumber
\end{equation}
Therefore,
\begin{equation}
\mathcal{N}_F(\rho_X)
=
F_Q\!\left(\rho_X,\frac{\sigma_z}{\sqrt2}\otimes\mathbb I_B\right).
\label{eq:qfi_sigma_z}
\end{equation}

The state $\rho_X$ decomposes into two invariant $2\times2$ blocks:
\begin{equation}
\rho_1=
\begin{pmatrix}
a & u\\
u^* & d
\end{pmatrix}
\quad \text{on } \mathrm{span}\{|00\rangle,|11\rangle\},
\qquad
\rho_2=
\begin{pmatrix}
b & v\\
v^* & c
\end{pmatrix}
\quad \text{on } \mathrm{span}\{|01\rangle,|10\rangle\}. \nonumber
\end{equation}
For each block, the generator has eigenvalues $\pm 1/\sqrt2$. Using the spectral formula of the quantum Fisher information, one obtains
\begin{equation}
F_Q(\rho_1)=\frac{8|u|^2}{a+d},
\qquad
F_Q(\rho_2)=\frac{8|v|^2}{b+c}.
\end{equation}
Hence,
\begin{equation}
\mathcal N_F(\rho_X)
=
F_Q(\rho_1)+F_Q(\rho_2)
=
8\left(
\frac{|u|^2}{a+d}
+
\frac{|v|^2}{b+c}
\right). \nonumber
\end{equation}
This completes the proof.
\end{proof}
\section{Noise resilience}
\label{Dynamics}
In realistic quantum-information processing tasks, quantum systems inevitably interact with their surrounding environment, leading to decoherence and degradation of quantum correlations. Therefore, it is important to investigate the robustness of the proposed QFI-based MIN (QFI-MIN) under noisy quantum channels. In this section, we study the dynamics of QFI-MIN under the amplitude damping channel, depolarizing noise, dephasing and genralized amplitude damping channel. Under the noisy channels, the time evolution of the  density matrix $\rho(0)$ is given by
\begin{equation}
\rho_{\gamma}(t)
=
\sum_{i=0}^{1}
(E_i \otimes I)
\rho(0)
(E_i^\dagger \otimes I).
\label{eq:ADCstate}
\end{equation}
where $E_i$ are the Kraus operators satisfy the completeness relation
\begin{equation}
\sum_i E_i^\dagger E_i = I. \nonumber
\end{equation}
\subsection{Amplitude Damping Channel}
The amplitude damping channel, which models irreversible energy dissipation processes such as spontaneous emission, is acting on subsystem $A$, described by the Kraus operators
\begin{equation}
E_0=
\begin{pmatrix}
1 & 0\\
0 & \sqrt{1-\gamma}
\end{pmatrix},
\qquad
E_1=
\begin{pmatrix}
0 & \sqrt{\gamma}\\
0 & 0
\end{pmatrix}, \nonumber
\label{eq:ADCkraus}
\end{equation}
where $0 \leq \gamma \leq 1$ denotes the decoherence strength. 
To illustrate the dynamics explicitly, we consider the maximally entangled Bell state
\begin{equation}
|\Phi^{+}\rangle
=
\frac{1}{\sqrt{2}}
\left(
|00\rangle + |11\rangle
\right),
\label{eq:Bellstate}
\end{equation}
whose corresponding density matrix evolves under the amplitude damping channel and non-matrix elements are $\rho_{11}=1/2,\rho_{22}=\gamma, \rho_{44}=1-\gamma$ and $\rho_{14}=\sqrt{1-\gamma}$. The evolved state retains the two-qubit $X$-state structure. One identifies
\begin{equation}
a=\frac{1}{2},
\quad
b=\frac{\gamma}{2},
\quad
c=0,
\quad
d=\frac{1-\gamma}{2},
\quad 
u=\frac{\sqrt{1-\gamma}}{2},
\qquad
v=0. \nonumber
\end{equation}
Using the analytical expression for QFI-MIN of two-qubit $X$ states, we obtain the QFI-based MIN under amplitude damping noise becomes 
\begin{equation}
\mathcal{N}_F(\rho_{\gamma})
=
\frac{4(1-\gamma)}{2-\gamma}.
\label{eq:QFIMINADC} 
\end{equation}

\begin{figure}%
  \centering
   \includegraphics[scale=0.4]{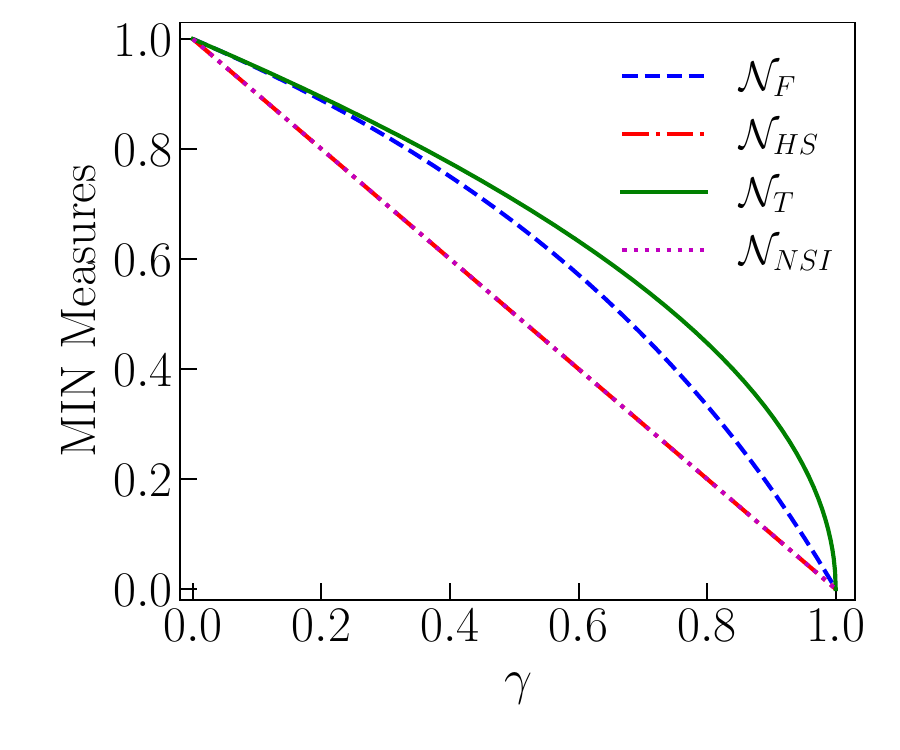}
    \caption{Comparison of QFI-MIN $N_F/2$ (dashed), Hilbert--Schmidt MIN $2N_{HS}$ (dot-dashed), trace MIN $N_T$ (solid), and skew-information MIN $N_{NSI}$ (dotted) as functions of the decoherence strength $\gamma$ under the amplitude damping channel. }
   \label{ADC}
\end{figure}

Equation~(\ref{eq:QFIMINADC}) shows that the QFI-based MIN decreases monotonically with increasing decoherence strength $\gamma$. In the absence of noise ($\gamma=0$), one obtains $\mathcal{N}_F(\rho_{0})=2$, corresponding to the maximal locally invariant quantum correlation of the Bell state. On the other hand, for complete damping $\gamma=1$, $\mathcal{N}_F(\rho_{1})=0$, indicating the complete destruction of Fisher-information-induced nonlocal correlations. Figure \ref{ADC} illustrates the behavior of different measurement-induced nonlocality measures under the amplitude damping channel as a function of the decoherence strength $\gamma$. It is observed that all measures decrease monotonically with increasing decoherence, indicating the progressive degradation of quantum correlations due to energy dissipation. For weak decoherence ($\gamma \ll 1$), the QFI-MIN behaves as
\begin{equation}
\mathcal{N}_F(\rho_{\gamma})
\approx
2-\gamma + O(\gamma^2), \nonumber
\end{equation}
showing an initial linear decay under amplitude damping noise.
The above analysis shows that amplitude damping suppresses the infinitesimal statistical distinguishability associated with locally invariant unitary encodings. Consequently, the QFI-based MIN decreases monotonically with increasing decoherence strength, reflecting the degradation of correlation-assisted metrological and communication resources due to irreversible energy dissipation. Nevertheless, the QFI-MIN remains nonzero for intermediate damping strengths, indicating partial robustness of Fisher-information-induced correlations against decoherence.

Unlike the original Hilbert--Schmidt MIN, the proposed QFI-MIN is invariant under local ancilla addition and possesses a direct operational interpretation in terms of quantum metrology, local channel discrimination, and correlation-assisted communication. These results demonstrate that QFI-MIN provides a more physically robust and operationally meaningful characterization of quantum correlations in noisy environments.
\begin{figure}%
  \centering
   \includegraphics[scale=0.4]{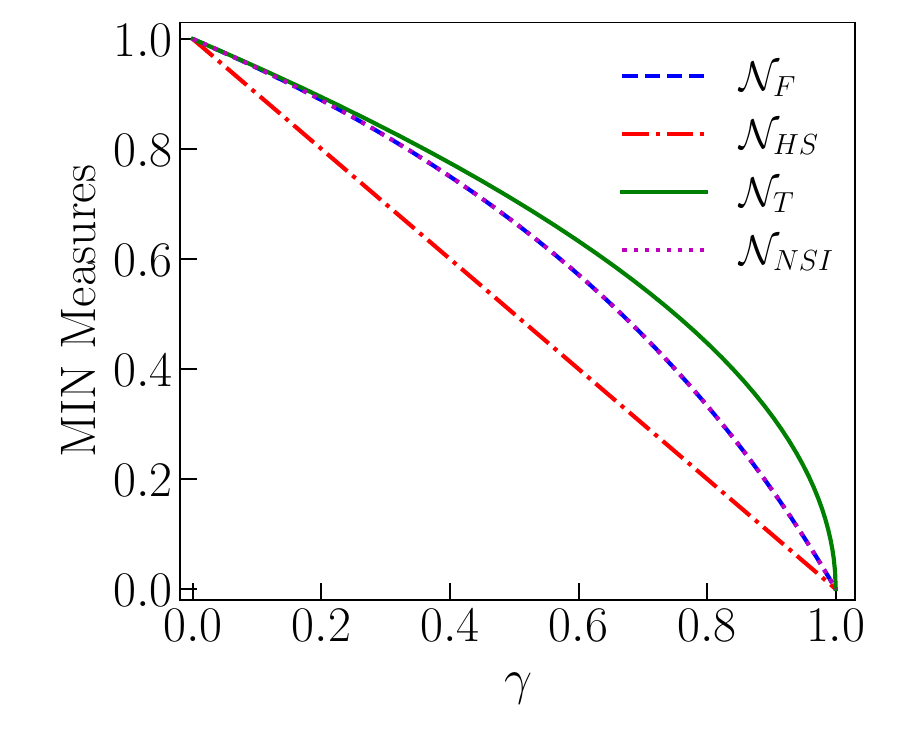}
    \caption{Evolution of QFI-MIN $N_F/2$ (dashed), Hilbert--Schmidt MIN $2N_{HS}$ (dot-dashed), trace MIN $N_T$ (solid), and skew-information MIN $N_{NSI}$ (dotted)  under the generalized amplitude damping channel as a function of $\gamma$}
   \label{GADC}
\end{figure}%

\begin{figure}%
  \centering
   \includegraphics[scale=0.4]{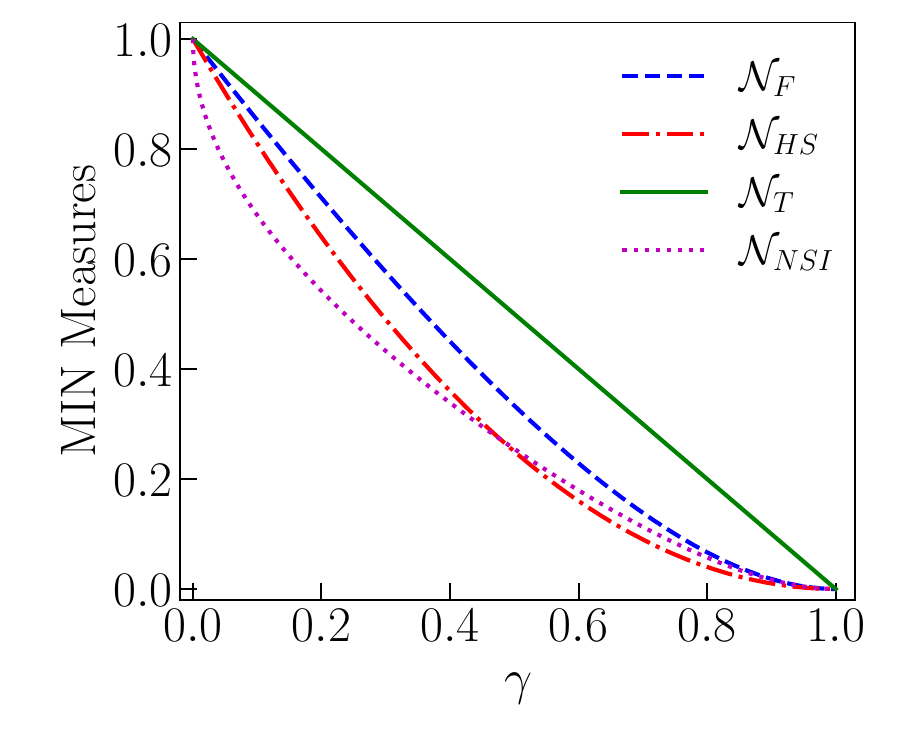}
    \caption{Behavior of QFI-MIN $N_F/2$ (dashed), Hilbert--Schmidt MIN $2N_{HS}$ (dot-dashed), trace MIN $N_T$ (solid), and skew-information MIN $N_{NSI}$ (dotted) versus the depolarizing strength $\gamma$.}
   \label{DEPOL}
\end{figure}%

\subsection{Generalized Amplitude Damping Noise}

We now investigate the behavior of the proposed QFI-based measurement-induced nonlocality (QFI-MIN) under the generalized amplitude damping (GAD) channel, which describes energy exchange between the quantum system and a thermal reservoir at finite temperature. The channel is characterized by the Kraus operators

\begin{equation}
E_0=\sqrt{p}
\begin{pmatrix}
1 & 0 \\
0 & \sqrt{1-\gamma}
\end{pmatrix},
\qquad
E_1=\sqrt{p}
\begin{pmatrix}
0 & \sqrt{\gamma} \\
0 & 0
\end{pmatrix}, \nonumber
\end{equation}

\begin{equation}
E_2=\sqrt{1-p}
\begin{pmatrix}
\sqrt{1-\gamma} & 0 \\
0 & 1
\end{pmatrix},
\qquad
E_3=\sqrt{1-p}
\begin{pmatrix}
0 & 0 \\
\sqrt{\gamma} & 0
\end{pmatrix}, \nonumber
\end{equation}
where $\gamma \in [0,1]$ denotes the decoherence strength and $p \in [0,1]$ represents the equilibrium population of the ground state of the thermal reservoir. Applying the Kraus operators of the GAD channel to the initial Bell state introduced above, the evolved density matrix retains the two-qubit X-state structure and the matrix elements are
\begin{align}
a&=\frac{1}{2}\Big[p+(1-p)(1-\gamma)\Big], \nonumber\\
b&=\frac{p\gamma}{2},\nonumber\\
c&=\frac{(1-p)\gamma}{2}, \nonumber\\
d&=\frac{1}{2}\Big[(1-p)+p(1-\gamma)\Big], \nonumber\\
u&=\frac{\sqrt{1-\gamma}}{2}, \nonumber
\end{align}
with $0\le \gamma \le 1$ denoting the decoherence strength and $0\le p\le 1$ characterizing the reservoir temperature. The QFI-MIN under the generalized amplitude damping channel becomes
\begin{equation}
\mathcal{N}_F(\rho_{\rm GAD})
=
\frac{4(1-\gamma)}
{2-\gamma}.
\end{equation}
Interestingly, the temperature parameter $p$ does not appear in the final expression, indicating that the QFI-MIN depends only on the decoherence strength $\gamma$. For weak decoherence $(\gamma\ll1)$,
\begin{equation}
\mathcal{N}_F(\rho_{\rm GAD})
\simeq
2-\gamma+\mathcal{O}(\gamma^2), \nonumber
\end{equation}
demonstrating an approximately linear initial decay. At $\gamma=0$, these measures attain their maximal values,
\begin{equation}
\mathcal{N}_F(\rho_0)=2,\qquad
\mathcal{N}_{\rm T}(\rho_0)=1,\qquad
\mathcal{N}_{SI}(\rho_0)=1,\qquad
\mathcal{N}_{\rm HS}(\rho_0)=\frac12, \nonumber
\end{equation}
whereas all of them vanish in the limit $\gamma=1$. Figure \ref{GADC} depicts the evolution of QFI-MIN and other MIN measures under the generalized amplitude damping channel. The results show that all correlation measures gradually decay with increasing $\gamma$, reflecting the detrimental influence of thermal relaxation on nonclassical correlations.

The comparison clearly demonstrates that the trace-distance MIN exhibits the slowest decay because it depends directly on the coherence amplitude. In contrast, the Hilbert--Schmidt MIN decreases most rapidly since it is proportional to the square of the coherence. The QFI-MIN remains significantly larger than the Hilbert--Schmidt MIN throughout the evolution and displays enhanced robustness against thermal relaxation. Since QFI-MIN quantifies the maximal statistical distinguishability associated with locally invariant unitary encodings, it retains a direct operational interpretation in quantum metrology, local channel discrimination, and correlation-assisted communication. Therefore, the present analysis indicates that QFI-MIN provides a more physically meaningful and robust characterization of quantum correlations in noisy environments than geometric formulations of measurement-induced nonlocality.

\subsection{Depolarizing Noise}

We next investigate the behavior of the proposed QFI-based measurement-induced nonlocality under the depolarizing channel. The depolarizing channel describes a process in which the quantum state loses information and is replaced by the maximally mixed state with a certain probability. The corresponding Kraus operators are given by

\begin{equation}
E_0=
\sqrt{1-\frac{3\gamma}{4}},I,
\qquad
E_1=
\sqrt{\frac{\gamma}{4}},\sigma_x, \nonumber
\end{equation}

\begin{equation}
E_2=
\sqrt{\frac{\gamma}{4}},\sigma_y,
\qquad
E_3=
\sqrt{\frac{\gamma}{4}},\sigma_z, \nonumber
\end{equation}
where $0\leq \gamma \leq 1$ denotes the depolarizing strength. Comparing with Eq.~(54), one identifies
\begin{align}
a&=d=\frac{1}{2}-\frac{\gamma}{4},\
b&=c=\frac{\gamma}{4},\
u&=\frac{1-\gamma}{2},\
v&=0. \nonumber
\end{align}
Substituting the above matrix elements into Eq.~(67), the QFI-based measurement-induced nonlocality becomes
\begin{equation}
\mathcal{N}_F(\rho_{\gamma})
=\frac{4(1-\gamma)^2}
{2-\gamma}.
\end{equation}

At $\gamma=0$, the Bell state possesses maximal quantum correlations,

\begin{equation}
\mathcal{N}_F^{A}(\rho_0)=2,
\qquad
\mathcal{N}_{\rm T}(\rho_0)=1,
\qquad
\mathcal{N}_{\rm SI}(\rho_0)=1,
\qquad
\mathcal{N}_{\rm HS}(\rho_0)=\frac12. \nonumber
\end{equation}
As $\gamma$ increases, all measures decrease monotonically and eventually vanish at $\gamma=1$,
\begin{equation}
\mathcal{N}_F(\rho_1)
= \mathcal{N}_{\rm T}(\rho_1)= \mathcal{N}_{\rm SI}(\rho_1)=\mathcal{N}_{\rm HS}(\rho_1)=
0. \nonumber
\end{equation}
Figure \ref{DEPOL} presents the dependence of various MIN measures on the depolarizing strength $\gamma$. As the depolarizing noise increases, all measures decrease monotonically and eventually vanish, demonstrating the complete suppression of quantum correlations in the maximally mixed state limit.

The depolarizing channel suppresses the off-diagonal coherence of the Bell state while simultaneously mixing the populations. As a result, the quantum correlations quantified by all measurement-induced nonlocality measures decay monotonically with increasing noise strength. Among the considered measures, the trace MIN exhibits the slowest decay due to its linear dependence on the surviving coherence. The Hilbert--Schmidt MIN decreases most rapidly because of its quadratic dependence on the coherence amplitude. The QFI-MIN remains significantly larger than the Hilbert--Schmidt MIN throughout the evolution and demonstrates enhanced robustness against depolarizing noise. Owing to its direct connection with statistical distinguishability and parameter-estimation sensitivity, QFI-MIN provides a more physically meaningful characterization of quantum correlations in noisy quantum systems.

The proposed QFI-MIN remains more robust throughout the evolution, highlighting its operational significance in noisy quantum environments."

\section{Conclusions}
\label{sec:6}
In this work, we introduced a quantum Fisher information based measurement-induced nonlocality (QFI-MIN) as a novel quantifier of quantum correlations. By combining the locally invariant framework of measurement-induced nonlocality with the information-geometric properties of quantum Fisher information, the proposed measure provides a physically consistent and operationally meaningful characterization of nonclassical correlations. We established its fundamental properties, including positivity, local unitary invariance, monotonicity under local CPTP maps, and invariance under the addition of uncorrelated local ancillas, thereby overcoming the local ancilla problem encountered in Hilbert–Schmidt norm-based MIN. Analytical expressions were derived for bipartite pure states, arbitrary two-qubit states, and two-qubit X states, revealing a direct connection between QFI-MIN and entanglement for pure states. 

Furthermore, we demonstrated that QFI-MIN possesses clear operational interpretations in local channel discrimination, quantum parameter estimation, and correlation-assisted classical communication, where it quantifies optimal distinguishability, achievable precision, and communication capability under locally invariant encodings. The investigation of noisy environments showed that QFI-MIN remains robust under amplitude damping, depolarizing, and generalized amplitude damping channels while faithfully capturing the degradation of quantum correlations induced by decoherence. 

Overall, the proposed framework establishes a unified link between quantum correlations, quantum metrology, and quantum information processing, and provides a promising foundation for exploring nonclassical resources in quantum technologies and open quantum systems.


\bibliography{references}

\end{document}